\documentclass[twocolumn,aps,prd,superscriptaddress,preprintnumbers,nofootinbib,floatfix]{revtex4}[11pt]
\pdfoutput=1
\usepackage{float}
\usepackage{amsmath,amssymb,graphicx} 
\graphicspath{{figs/}}
\usepackage{epsfig,verbatim}
\usepackage{hyperref}
\usepackage{comment}
\usepackage{color}
\usepackage{slashed}
\usepackage{subfigure}
\usepackage[usenames,dvipsnames]{xcolor}
\usepackage{comment}
\usepackage{mdwlist, paralist}
\usepackage{dcolumn}
\usepackage{bm}
\usepackage{subfigure}

\input babarsym.tex

\def\to{\rightarrow}
\definecolor{darkpastelgreen}{rgb}{0.01, 0.75, 0.24}

\newcommand{\simgt}{\mathrel{\lower2.5pt\vbox{\lineskip=0pt\baselineskip=0pt
           \hbox{$>$}\hbox{$\sim$}}}}
\newcommand{\simlt}{\mathrel{\lower2.5pt\vbox{\lineskip=0pt\baselineskip=0pt
           \hbox{$<$}\hbox{$\sim$}}}}

\newcommand{\beq}{\begin{equation}}
\newcommand{\eeq}{\end{equation}}
\newcommand{\bea}{\begin{eqnarray}}
\newcommand{\eea}{\end{eqnarray}}
\newcommand{\bit}{\begin{itemize}}
\newcommand{\eit}{\end{itemize}}

\newcommand{\eqssref}[3]{Eqs.~(\ref{eq:#1}), (\ref{eq:#2}) and (\ref{eq:#3})}
\newcommand{\Eqref}[1]{Eq.~(\ref{eq:#1})}

\newcommand{\ie}{{\it i.e.},}
\newcommand{\eg}{{\it e.g.}}

\newcommand{\mue}{Mu3e}

\begin{document} 

\title{Projections for Dark Photon Searches at Mu3e}

\preprint{YITP-SB-14-36}

\author{Bertrand Echenard}
\thanks{echenard@hep.caltech.edu}
\affiliation{California Institute of Technology, Pasadena, California 91125}

\author{Rouven Essig}
\thanks{rouven.essig@stonybrook.edu}
\affiliation{C.N.~Yang Institute for Theoretical Physics, Stony Brook University, Stony Brook, NY 11794}

\author{Yi-Ming Zhong}
\thanks{yiming.zhong@stonybrook.edu}
\affiliation{C.N.~Yang Institute for Theoretical Physics, Stony Brook University, Stony Brook, NY 11794}

\begin{abstract} 
We show that dark photons ($A'$) with masses $\sim 10-80 \mev$ can be probed in the decay $\mu^+\to e^+ \nu_e \bar{\nu}_\mu A'$, 
$A' \to e^+e^-$, with the upcoming \mue\ experiment at the Paul Scherrer Institute (PSI) in Switzerland.  
With an expected $10^{15}$ ($5.5\times 10^{16}$) muon decays in 2015--2016 (2018 and beyond), \mue\ 
has the exciting opportunity to probe a substantial fraction of currently unexplored dark photon parameter space, 
probing kinetic-mixing parameter, $\epsilon$,  as low as $\epsilon^2 \sim 10^{-7}~(10^{-8})$. 
No modifications of the existing \mue\ setup are required.  
\end{abstract}


\maketitle

\setcounter{equation}{0} \setcounter{footnote}{0}

\section{INTRODUCTION}
\label{sec:intro}

\begin{figure}[]
  \centering
  \includegraphics[width=0.3 \textwidth]{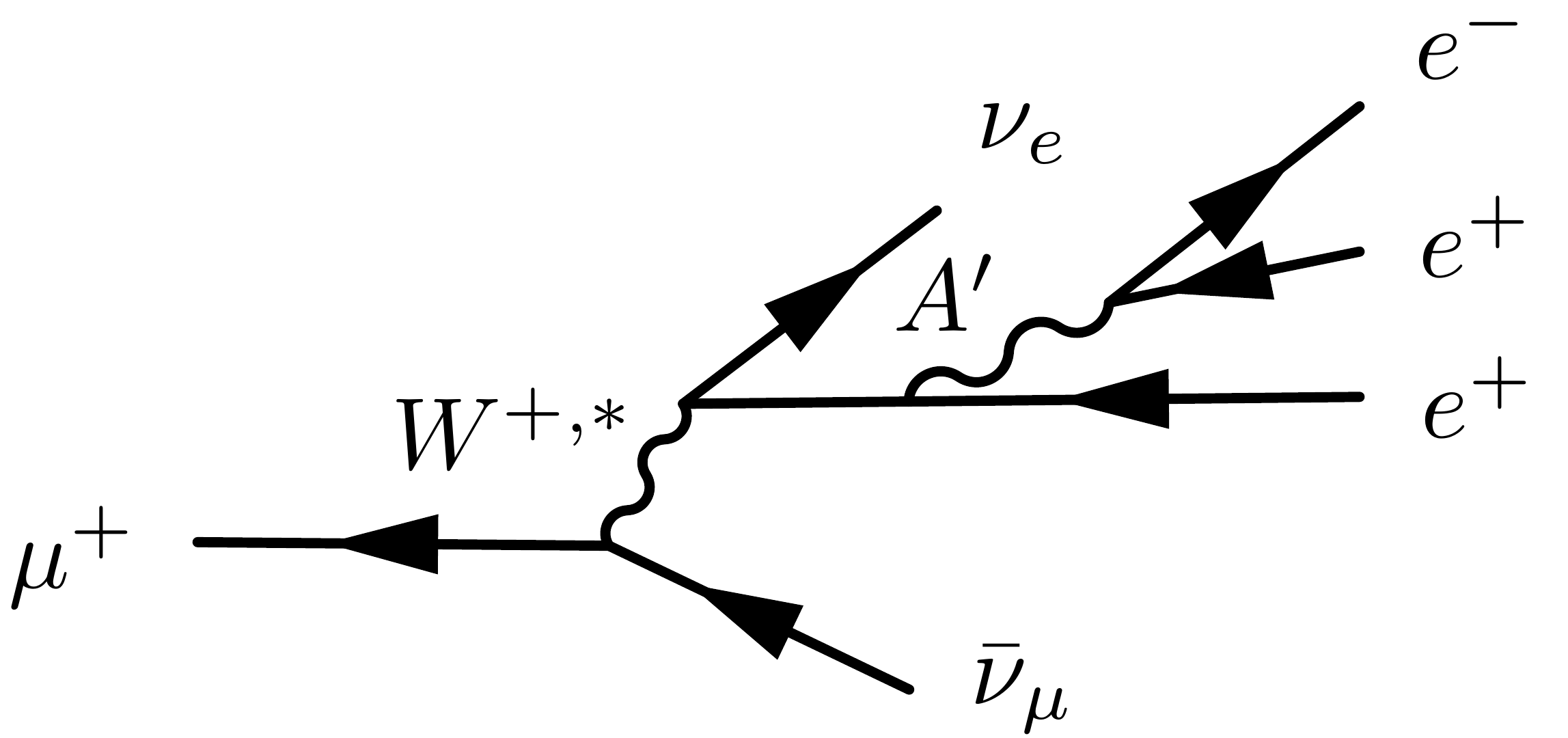}\hspace*{0.5em}\\
\vskip 5mm
  \includegraphics[width=0.3 \textwidth]{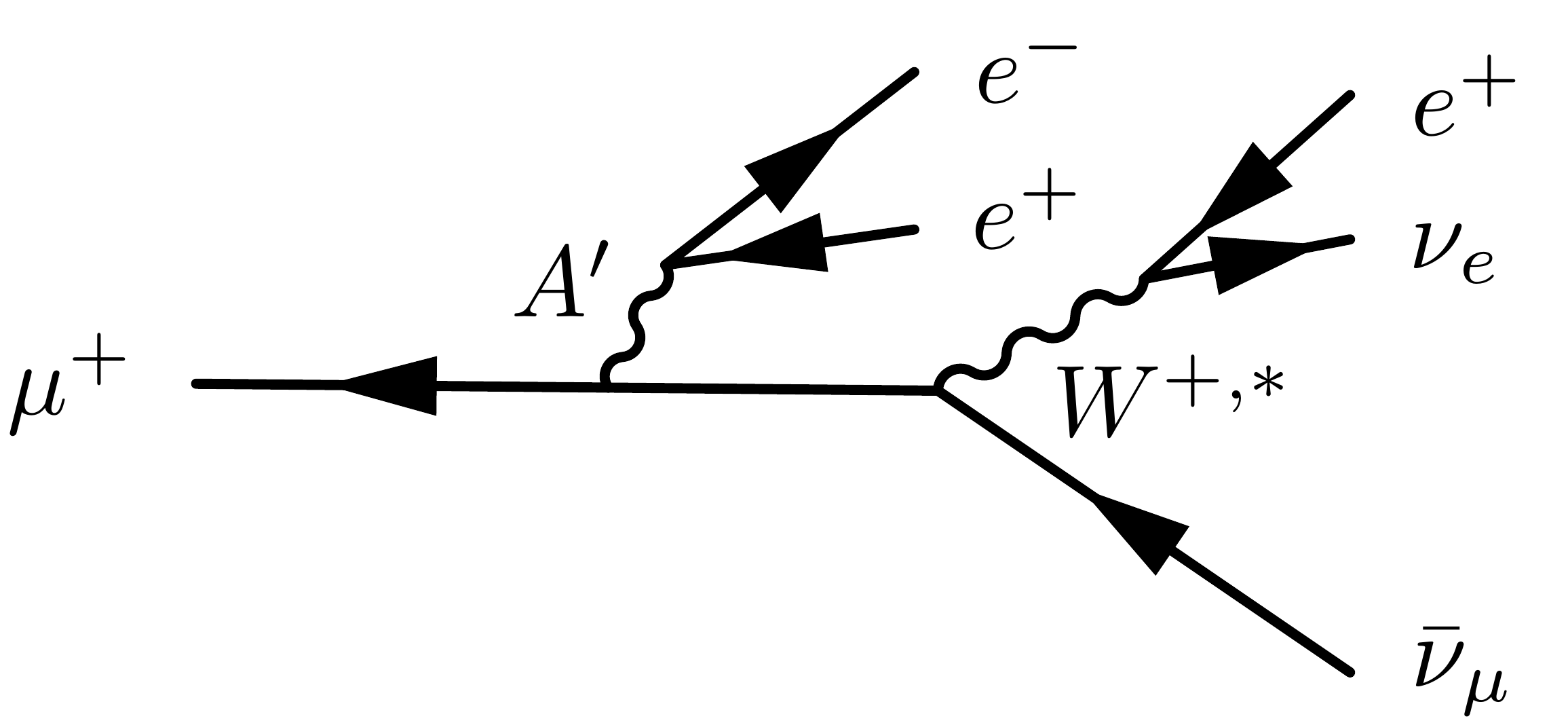}\hspace*{0.5em}\\
\vskip 5mm
    \includegraphics[width=0.3 \textwidth]{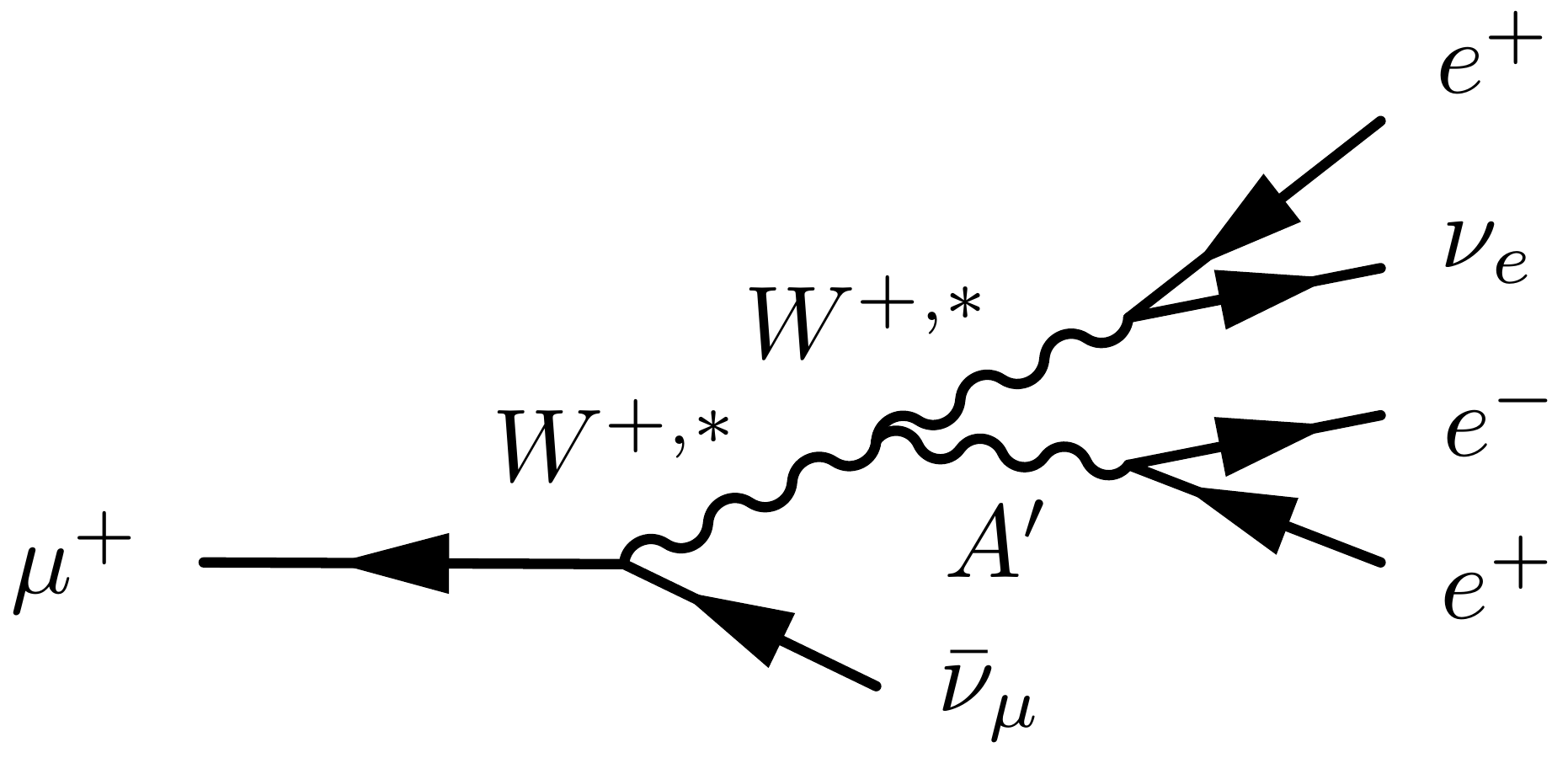}\hspace*{0.5em}\\
  \caption{Feynman diagrams for (on-shell) dark photon production in muon decays, 
  $\mu^+\to e^+ \nu_e \bar \nu_\mu A',~A' \to e^-e^+$.}
  \label{fig:feyndiagrams}
\end{figure}

There are only a few ways in which new particles and forces below the weak-scale can interact with the standard 
model (SM) particles and have remained undetected thus far. Among the simplest possibilities is the existence 
of a light, massive vector boson called a dark photon ($A'$). A substantial effort is underway to search for a 
dark photon with a variety of experiments. In this paper, we show that the upcoming Mu3e experiment at the Paul 
Scherrer Institute (PSI) in Switzerland is also sensitive to dark photons. Using an unprecedented number of 
muon decays\footnote{``Muon" refers to $\mu^+$ in this paper.} 
in their search for the lepton flavor violating decay $\mu^+\!\to\! e^+e^-e^+$, 
Mu3e can also search for the decay $\mu^+\!\to\! e^+ \nu_e \bar \nu_\mu A',~A'\! \to\! e^+e^-$ 
shown in Fig.~\ref{fig:feyndiagrams}.  
This allows them to probe currently unexplored regions of the dark photon parameter space.  
We note that while our focus will be on vector bosons (the dark photon), other particles 
that couple to electrons and/or muons and decay to an $e^+e^-$ pair could also be probed with Mu3e.  
 
The dark photon is the mediator of a new, broken U(1)$_{\rm D}$ gauge group and appears in many theoretical scenarios, see 
\eg~\cite{Essig:2013lka,Hewett:2012ns,Jaeckel:2010ni} and references therein.  
It can interact with ordinary matter through ``kinetic mixing''~\cite{Holdom:1985ag,Galison:1983pa,Dienes:1996zr} with the 
SM hypercharge, U(1)$_\text{Y}$, gauge boson.  
At low energies, the dominant effect is  a mixing of the U(1)$_{\rm D}$ with the SM photon, $\rm U(1)_{\rm EM}$, 
as described with the Lagrangian 
\beq
\mathcal{L} = \mathcal{L}_{\rm SM} - \frac{\epsilon}{2} F'_{\mu\nu}F^{\mu\nu} - \frac{1}{4} F'_{\mu\nu} F'^{\mu\nu} + \frac{1}{2} m_{A'}^2 A'_\mu  A'^{\mu}\,.
\eeq  
Here $\mathcal{L_{\rm SM}}$ is the SM Lagrangian, $\epsilon$ is the kinetic mixing parameter, $F'^{\mu\nu}$ ($F^{\mu\nu}$) is the 
U(1)$_{\rm D}$ (U(1)$_{\text{EM}}$) field strength, and $m_{A'}$ is the dark photon mass (the mechanism for generating this mass 
is not important for our purposes). The mixing between the dark photon and the SM photon leads to an $\epsilon$-suppressed coupling 
of the dark photon to the electromagnetic current, $J^\mu_{\rm EM}$, \ie~to quarks and charged leptons,
\beq\label{eq:mixing}
\mathcal{L} \supset \epsilon \, e \, A'_\mu\, J_{\rm EM}^\mu\,.
\eeq
The two relevant parameters of the model are the kinetic mixing parameter and the dark photon mass.  
The coupling in Eq.~(\ref{eq:mixing}) allows the dark photon to be probed with a wide range of experiments, 
see \eg~\cite{Essig:2013lka,Hewett:2012ns,Jaeckel:2010ni} for a recent review and references.  
We do not consider the addition of other low-mass particles to this model.

Theoretically, the values of the kinetic mixing and the dark photon mass can take on a wide range of values.  
However, much attention has recently been focused on the MeV--GeV mass range.  
In this mass range, the dark photon could explain the $\sim 3.6\sigma$ discrepancy between the observed and SM value of 
the muon anomalous magnetic moment ($a_\mu \equiv g_\mu-2$, where $g_\mu$ is the muon's gyromagnetic ratio)~\cite{Pospelov:2008zw,Bennett:2006fi,Davier:2010nc} and offer an explanation for 
various dark matter related anomalies through dark matter-dark photon interactions~\cite{ArkaniHamed:2008qn,Pospelov:2008jd,Finkbeiner:2007kk,Fayet:2004bw}.  
Moreover, a dark photon mass  in this range can be generated naturally in several new physics scenarios~\cite{ArkaniHamed:2008qp,
Cheung:2009fk,Baumgart:2009tn,Morrissey:2009ur,Essig:2009nc}.  
In addition, if U(1)$_\text{Y}$ is embedded in a Grand Unified Theory (GUT), 
the mixing can be generated by a one-(two-)loop interaction and naturally give 
$\epsilon \sim 10^{-3}-10^{-1}$ ($\sim 10^{-5}-10^{-3}$)~\cite{ArkaniHamed:2008qp,Holdom:1985ag,Essig:2010ye,Baumgart:2009tn}.  

There are many experimental probes of MeV--GeV mass dark photons that decay directly to SM particles.  
These include collider experiments, beam dumps, rare meson decays, supernova cooling, and precision measurements~\cite{
Bjorken:2009mm,Bjorken:1988as,Riordan:1987aw,Bross:1989mp,Batell:2009yf,Strassler:2006im,Strassler:2006qa,Essig:2009nc,Freytsis:2009bh,Essig:2010xa,Blumlein:2011mv,Andreas:2012mt,Pospelov:2008zw,Reece:2009un,Aubert:2009cp,Hook:2010tw,Babusci:2012cr,Archilli:2011zc,Abrahamyan:2011gv,Merkel:2014avp,Dent:2012mx,Davoudiasl:2012ig,Davoudiasl:2012ag, Davoudiasl:2013aya,Endo:2012hp,Balewski:2013oza,Adlarson:2013eza,Agakishiev:2013fwl,Andreas:2013lya,Battaglieri:2014hga,Lees:2014xha,Adare:2014ega,Kazanas:2014mca,Blumlein:2013cua,CERNNA48/2:2015lha}.    
Existing constraints have almost disfavored the entire mass and coupling range in which dark photons could  
explain the mismatch between the observed and SM expected value of $a_\mu$, 
assuming the dark photon decays directly to SM particles with a branching ratio close to 100\%.  
A reduced branching ratio is possible if there exist other light particles that couple to the dark photon and open up 
additional decay modes. 

In this paper, we will show that Mu3e 
can probe dark photons in the mass range $2 m_e < m_{A'} < m_{\mu}$, where $m_e$ ($m_\mu$) is the electron (muon) mass, 
and improve upon current constraints on $\epsilon$ in the range 
$10~{\rm{MeV}} \lesssim m_{A'} \lesssim 80~{\rm MeV}$, down to $\epsilon^2 \sim 10^{-8}$. 
This probes well into the mentioned above parameter region motivated from embedding the U(1)$_\text{Y}$ in a GUT,
as well as probing $a_\mu$ favored dark photon to SM branching ratios significantly less than 100\%.   
Depending on the performance of the detector, \mue\ may also be sensitive to long-lived dark photons, which produce 
displaced vertices. 

The paper is organized as follows. 
We first summarize in Sec.~\ref{sec:mu3eSearch} how dark photons can be produced in muon decays and detected in Mu3e.  
We then discuss the projections for the sensitivity of dark photon searches at Mu3e for 
prompt decays in Sec.~\ref{sec:exclusionregion} and briefly discuss a possible search using displaced vertices in 
Sec.~\ref{sec:vertexregion}.  Our conclusions are presented in Sec.~\ref{sec:conclusions}.
In Appendix~\ref{sec:pu}, we detail our estimate of the accidental backgrounds for the prompt search.  

\section{The Search for Dark Photons With Mu3e}
\label{sec:mu3eSearch}

The Mu3e experiment at PSI \cite{Blondel:2013ia} has been proposed to search for the charged lepton flavor violating decay 
$\mu^+ \!\rightarrow \!e^+e^-e^+$ with an ultimate sensitivity of  $10^{-16}$, four orders below the current limits. 
It will take advantage of one of the most intense sources of muons in the world. 
During its first phase (2015 -- 2016), Mu3e will probe $10^{15}$ muon decays, 
and more than $5.5\times 10^{16}$ muon decays by the end of phase~II (2018 and beyond). 
To achieve the required sensitivity, a novel design based on high-granularity 
thin silicon pixel detectors, supplemented by a fast timing system, has been proposed. 

\begin{figure}[]
  \includegraphics[width=0.48 \textwidth]{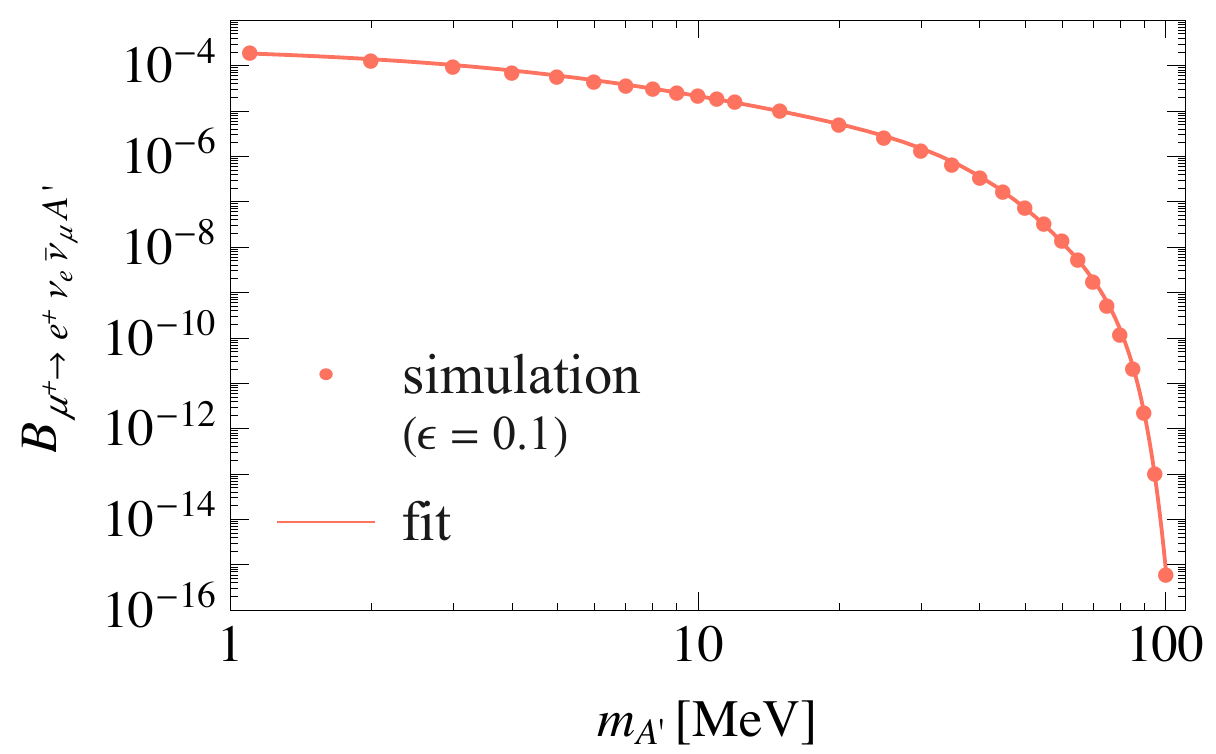} 
  \caption{The branching ratio of the muon decay channel $\mu^+\! \to \!e^+ \nu_e \bar \nu_\mu A'$ with $\epsilon=0.1$. 
  Shown are the numerical values computed with MadGraph5\_aMC@NLO (red points) and a parametrized fit to 
  these numerical values (red solid line), which is given by Eq.~(\ref{eq:decaywidth}).
  \label{fig:decaywidth}
    }
\end{figure}

The large statistics and excellent detector resolution offer an ideal setup to search for dark photon production in muon 
decays as well. The production mechanism is illustrated in 
Fig.~\ref{fig:feyndiagrams}: the dark photon can be either emitted from the initial state radiation off the $\mu^+$, or final 
state radiation off the $e^+$, or radiate off the internal $W$-boson.  
The latter process is suppressed by $\sim m_\mu^2/m_W^2 \sim 10^{-6}$ at the amplitude level 
compared to the other processes due to the different propagators appearing in the diagrams 
(this is similar for the corresponding SM process where the dark photon is replaced by the SM photon, see also~\cite{Djilkibaev:2008jy}).  
The corresponding decay width of $\mu^+ \rightarrow e^+ \nu_e \bar\nu_\mu A'$, 
is evaluated using MadGraph5\_aMC@NLO 2.1.0~\cite{Alwall:2014hca} for $m_{A'}$ ranging 
from $1.1 \mev$ to $100 \mev$. 
Approximating the total decay width, $\Gamma_{\rm tot}$, as the SM muon decay width, the resulting branching ratio $B_{\text{sig}}$ 
is presented in Fig.~\ref{fig:decaywidth} for $\epsilon=0.1$.  
We also include a parametrized curve (red, labeled ``fit'') of the form 
$B_{\text{sig}}=B_{\text{sig}}(\epsilon, m_{A'})$ 
with  
\beq
B_{\text{sig}}=\frac{1}{3\times 10^{-19}}\left(\frac{\epsilon}{0.1}\right)^2\exp\left(\sum_{i=0}^5 a_i \left(\frac{m_{A'}}{\gev}\right)^i \right),
\label{eq:decaywidth}
\eeq
where $a_0=-50.866$, $a_1=-360.93$, $a_2 =13998.59$,  $a_3=-3.731\times 10^5$, $a_4= 4.442\times 10^6$, 
$a_5= -2.015\times 10^7$, and we take the fine structure constant $\alpha=1/137.036$ 
and  $\Gamma_{\rm tot} \simeq 3\times 10^{-19} \gev$~\cite{pdg}. 

\begin{figure*}[t!]
 \centering
 \includegraphics[width=0.9\textwidth]{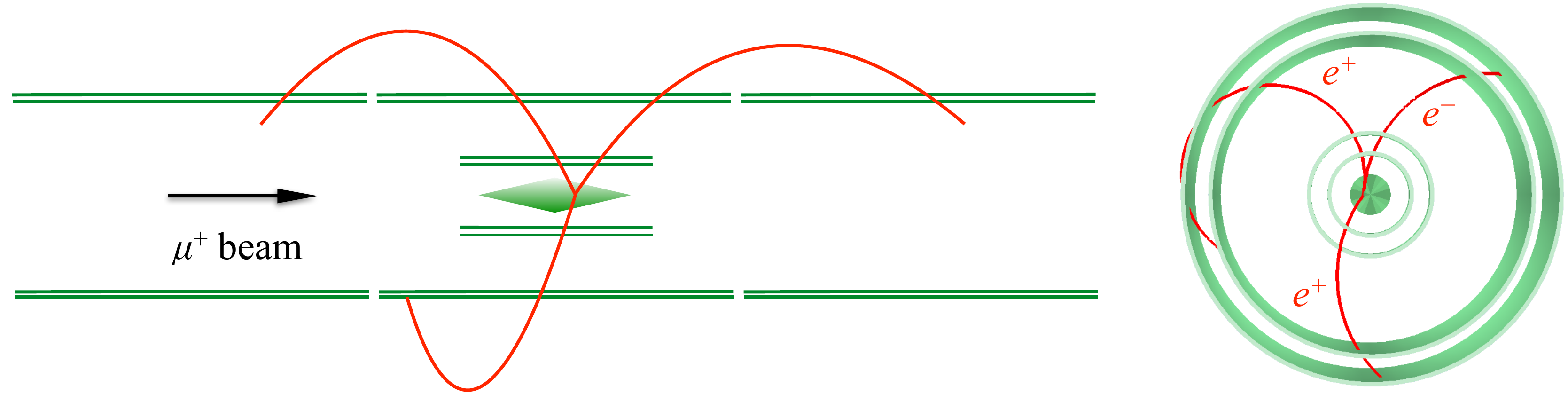}
\caption{
{\bf Left:} Side view of the experimental setup (adapted from~\cite{Blondel:2013ia}).  
A muon beam impinges on a target consisting of two hollow aluminum cones connected at their base. 
A silicon tracker composed of two inner and two outer double layers of cylindrical pixel silicon 
surrounds the target. Although not included in the simulation, a time-of-flight device provides 
a timing measurement with a resolution of 250~ps. A simulated $\mu^+ \rightarrow e^+ \nu_e \bar\nu_\mu A', ~A' \rightarrow e^+e^-$ 
event is shown. {\bf Right:} Transverse view of the experimental setup.  The stopping target is shown at the center, surrounded 
by two inner and two outer cylindrical layers of silicon detectors. 
}
\label{fig::detector}
\end{figure*}

For $2 m_e \le m_{A'} \le 2 m_\mu$, the dominant decay is $A' \to e^+e^-$ (the loop-induced decay $A' \to 3 \gamma$ is highly 
suppressed and only important for $m_{A'} < 2 m_e$). The main signature of such a dark photon is that the invariant mass of the 
$e^+e^-$-pair from the dark photon decay must equal $m_{A'}$. The invariant mass spectrum is dominated by SM background events 
(see Sec.~\ref{sec:mu3eprojection}), but a resonance search or ``bump-hunt'' can be used to search for a dark photon.  

In addition to a resonance search, displaced $e^+e^-$ vertices can also be used to probe long-lived dark photons.  
The dark photon width and the proper decay length 
are given by, respectively, 
\bea
\Gamma_{A' \to e^+e^-} &=&\frac{\alpha\, \epsilon^2}{3}\, m_{A'}\, \sqrt{1-\frac{4 m_e^2}{m_{A'}^2}} \,\left(1+\frac{2 m_e^2}{m_{A'}^2}\right),\,\\
c \,\tau_{A'\to e^+e^-} & \simeq & 0.8~{\rm mm}\, \left(\frac{10^{-4}}{\epsilon}\right)^2 \, \frac{10~{\rm MeV}}{m_{A'}}\,.
\label{eq:width}
\eea
For small-enough values of $\epsilon$, 
the dark photon will travel a finite distance and the $e^+e^-$-pair will be reconstructed as a displaced vertex (for even smaller 
values, the decay length will be large enough to allow for the shielding of almost any backgrounds, as in beam-dump 
experiments). Since the backgrounds are expected to be greatly reduced with respect to prompt decays, displaced vertices could 
provide sensitivity to low values of the kinematic mixing.

\section{Projections for Dark Photon Searches With Mu3e}
\label{sec:mu3eprojection}

The sensitivity to dark photons with an experimental setup similar to that of \mue\ is 
studied using a simulation program, FastSim, that was originally developed for the 
Super$B$ experiment~\cite{Baszczyk:2013xua}, based on the software framework and analysis tools used 
by the \babar\ collaboration~\cite{Aubert:2001tu,TheBABAR:2013jta}. 
Detector components are described in FastSim as two-dimensional shells of 
geometric objects, such as cylinders, disks, or planes, and the effect of the physical thickness is modeled parametrically. Coulomb scattering and 
energy loss by ionization are described with the standard parametrization in terms of radiation length 
and particle momentum. 
Simplified cross sections are used to describe Bremsstrahlung and pair production. 
Tracking measurements are simulated in terms of single-hit and two-hit resolutions, 
while silicon strip detectors are modeled as two independent orthogonal projections. 
Tracks are reconstructed from the simulated hits passed to the \babar\ Kalman filter track fitting algorithm. 
Uncertainties associated with pattern recognition algorithms traditionally used to form track hits are 
introduced using models based on the \babar\ pattern recognition algorithm performance. 

The FastSim model is a simplified version of the proposed Mu3e detector~\cite{Blondel:2013ia}, which consists  
of a silicon tracker composed of two inner and two outer double layers of cylindrical pixel silicon detectors surrounding 
the target. The inner layers have a length of 12~cm, while the outer silicon layers are extended to a 
length of 180~cm to improve the momentum resolution of recurling tracks. The innermost (outermost) silicon 
detectors are placed at a radius of 1.9~cm (8.9~cm). Silicon sensors are simulated as 50~$\mu$m thick double-sided striplet sensors mounted on 50~$\mu$m of kapton in FastSim. The spatial resolution of the hits is modeled as a sum of two components with resolutions of 8~$\mu$m and 20~$\mu$m. Although Mu3e uses pixel silicon sensors, we expect the performances of both tracking system to be comparable.
The target is composed of two hollow aluminum cones connected at their base. 
Each cone is 5~cm long, 50~$\mu$m thick with a base radius of 1~cm. 
The entire detector is placed in a 1~T solenoidal magnetic field. 
Although not included in FastSim, a time-of-flight device provides a timing measurement. 
We assume a time resolution of 250~ps, averaging the values of the corresponding Mu3e detector systems. 
We define a coordinate system having the $z$-axis aligned along the axis of the cylindrical silicon detectors, 
with the transverse plane oriented perpendicular to the $z$-axis. 
The apparatus layout is displayed in Fig.~\ref{fig::detector}, together with a simulated 
$\mu^+ \rightarrow e^+ \nu_e \bar\nu_\mu A',~A' \rightarrow e^+e^-$ event.

\subsection{Promptly Decaying Dark Photons}
\label{sec:exclusionregion}

We begin by studying the sensitivity of prompt dark photon decays in 
$\mu^+ \rightarrow e^+ \nu_e \bar\nu_\mu A',~A' \rightarrow e^+e^-$ 
events. Large samples of signal and background events are generated to study the signal efficiency and background levels.  
We assume that muons decay uniformly at rest in the target. 
Signal events are generated with MadGraph5\_aMC@NLO 2.1.0 for $5 \mev < m_{A'} < 100 \mev$. 
The background processes can be classified as either {\em irreducible} or {\em accidental}: 
\bit
\item {\it Irreducible backgrounds} arise from events with internal conversions of the photon in 
$\mu^+ \rightarrow e^+ \nu_e \bar\nu_\mu \gamma^*(\rightarrow e^+e^-)$ decays, or 
from radiative muon decays where the radiated photon converts into an electron-positron pair inside the target material. 
Conversion outside the target material, \ie~in the detector material, can be efficiently tagged and are not considered. 
These background processes are simulated using the matrix element and differential decay width given 
in~\cite{Kuno:1999jp, Djilkibaev:2008jy}, and the events are normalized using the following branching fractions: 
$B_{\mu^+ \rightarrow e^+ \nu_e\bar\nu_\mu  e^+e^-}= (3.4 \pm 0.4)\times 10^{-5}$ and 
$B_{\mu^+ \rightarrow e^+ \nu_e \bar\nu_\mu \gamma}= (1.4 \pm 0.4)\%$~\cite{pdg}.  
As the probability of photon conversion inside the target is of ${\cal O}(10^{-3})$, both channels contribute 
roughly equally to the irreducible background.  

\begin{figure}[]
\begin{center}
\includegraphics[width=0.45\textwidth]{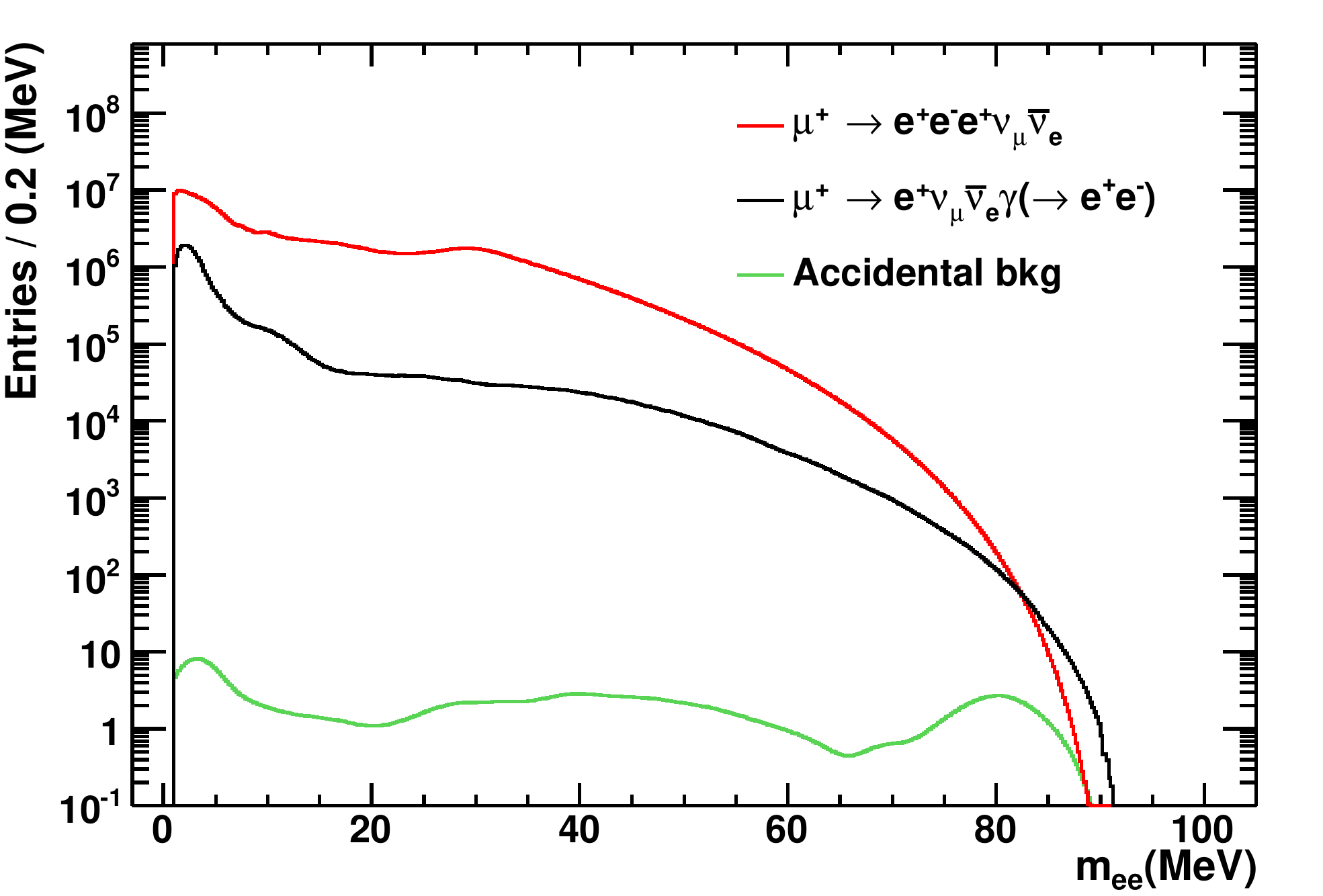}
\includegraphics[width=0.45\textwidth]{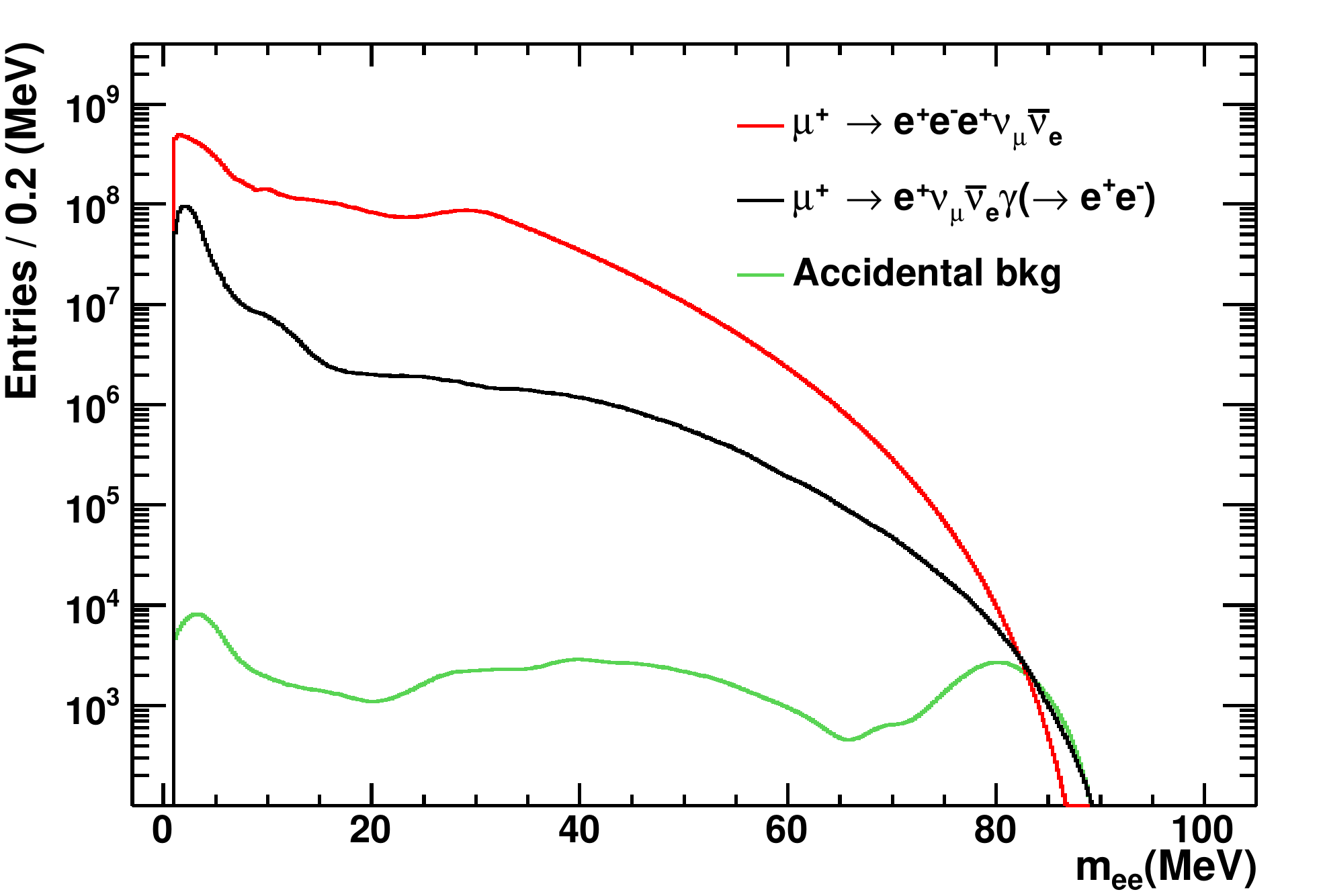}
\end{center}
\caption{The expected $\epem$ invariant mass distribution from the various sources of background displayed assuming 
a total number  of $10^{15}$ ({\bf top}) or $5.5\times 10^{16}$ ({\bf bottom}) muon decays for the phase I and II of Mu3e, respectively.
The accidental backgrounds include the 3M, 2M$_{\gamma}$, and 2M$_{3e}$ backgrounds but not the 2M$_{\rm Bhabha}$
background (see text for details).}
\label{Fig::eemass}
\end{figure}

\item {\it Accidental backgrounds} arise mainly from the combination of several muon decays where, \eg, one of the positrons is 
misreconstructed as an electron. 
We consider background sources from the following accidental combinations: 
(1) three Michel decays ($\mu^+ \rightarrow e^+ \nu_e \bar\nu_\mu$) where one positron is misreconstructed as an electron (``3M decays"), 
(2) a Michel decay and a radiative Michel decay ($\mu^+ \rightarrow e^+ \nu_e \bar\nu_\mu \gamma$) where the photon converts to a $e^+e^-$ pair in the target material and one positron remains undetected (``2M$_\gamma$ decays"), 
and (3) a Michel decay and a radiative Michel decay with internal conversion where one positron again remains undetected 
(``2M$_{3e}$ decays"). 
Another source of accidental background, which we will not include, 
arises from two Michel decays where the outgoing positron from one of the Michel 
decays produces an electron by interacting with the detector material via Bhabha scattering 
(``2M$_{\rm Bhabha}$ decays").
Other sources of accidental backgrounds are expected to be smaller.  
\eit

We generate signal and background events and process them with FastSim to determine the 
detection efficiency and the invariant mass distribution, $m_{e^+ e^-}$.  
We require all electrons and positrons to have a minimum transverse momentum of 10~MeV to match the \mue\ tracker acceptance.  
The $\mu^+ \rightarrow e^+ \nu_e\bar\nu_\mu  e^+e^-$ candidates are formed  
by combining two positrons and an electron, and fit with the constraint that the tracks originate from the same position at 
the surface of the target. 
We select only well reconstructed candidates by requiring the probability of the $\chi^2$ of the constrained fit to be 
greater than 1\%.  
Additional kinematic constraints can further distinguish $\mu^+ \rightarrow e^+ \nu_e\bar\nu_\mu  e^+e^-$ decays 
from accidental backgrounds. 
The magnitude of the sum of the momenta of the electron and two positrons 
($ |\vec{p}_{3e}|\equiv|\vec{p}_{e^-}+\vec{p}_{e^+,1}+\vec{p}_{e^+,2}|$) 
must be compatible with the muon decay hypothesis, requiring
\beq 
|\vec{p}_{3e}| \leq \frac{m_\mu^2 - m_{3e}^2}{2 m_\mu}, 
\label{eq:cut}
\eeq
where $m_{3e}$ is the invariant mass of the three tracks.  

\begin{figure}[!t]
\centering
\includegraphics[width=0.48 \textwidth]{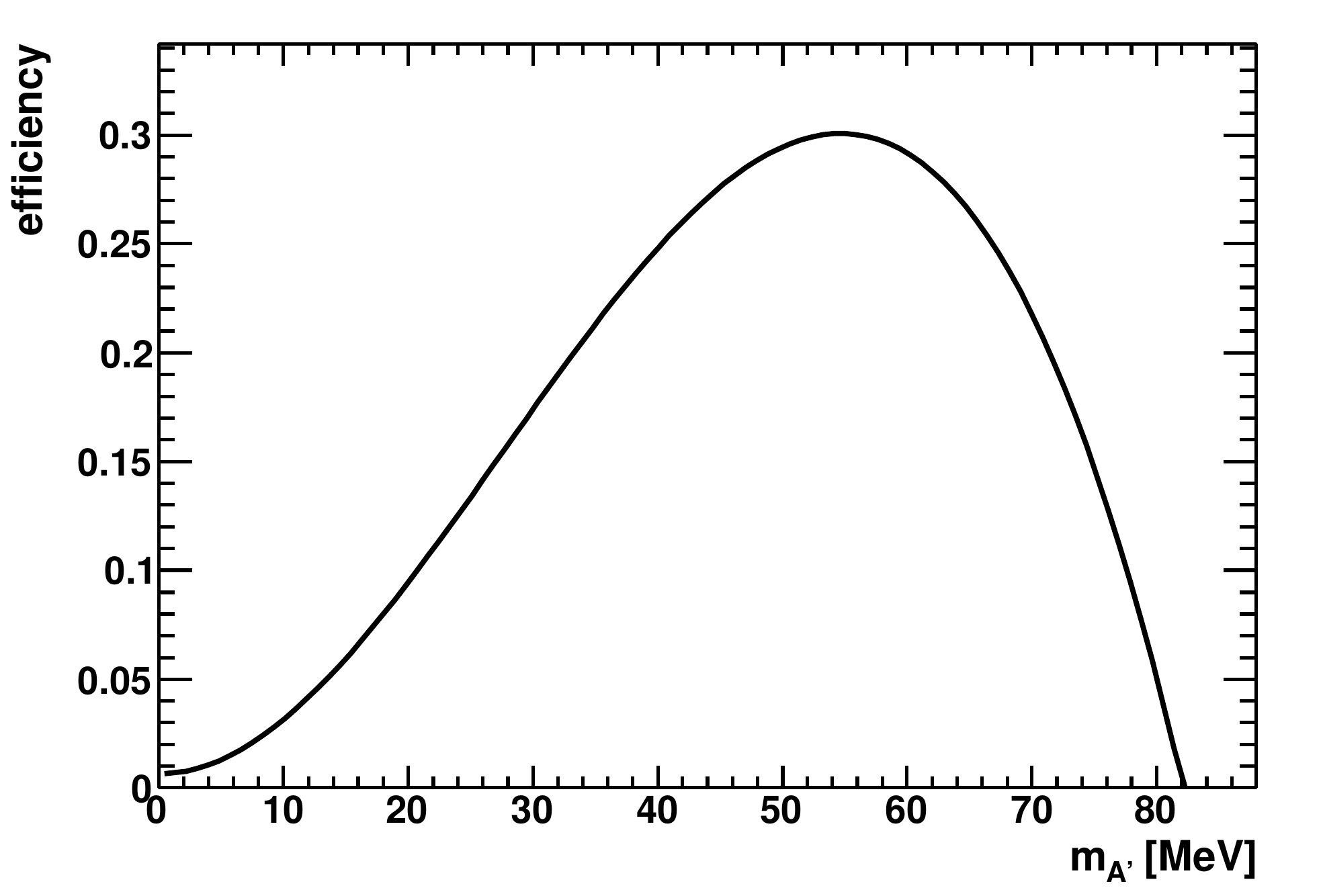}
\caption{The signal efficiency as a function of the dark photon mass ($m_{A'}$) for prompt $\mu^+ \rightarrow e^+ \nu_e \bar\nu_\mu A', ~A' \rightarrow e^+e^-$ decays.}
\label{fig:effiPrompt}
\end{figure}

While we can reliably determine the efficiency of the signal and irreducible backgrounds, it is more challenging to estimate 
the accidental backgrounds with very high accuracy.  
However, the following approach, which is similar to that described by Mu3e~\cite{Blondel:2013ia}, is sufficient for our purposes.  
The number of 3M, 2M$_{\gamma}$, and 2M$_{3e}$ decays are given by, respectively, 
\bea
N_{\rm 3M}\!&\simeq&\! \frac{1}{2}  N_\mu \, R_\mu^2 \,\delta t^2 \,P_\text{p}^2 \,B_{\mu^+ \rightarrow e^+ \nu_e \bar\nu_\mu}^3 \, P_{e^+\to e^-}\,, 
\label{eq:3M}\\
N_{\rm{2M}_\gamma} \!&\simeq&\! N_\mu \, R_\mu \,\delta t \, P_\text{p} \, B_{\mu^+ \rightarrow e^+ \nu_e \bar\nu_\mu}\,  B_{\mu^+ \rightarrow e^+ \nu_e \bar\nu_\mu \gamma}  \,P_\gamma\,,~~
\label{eq:2My}\\
N_{\text{2M}_{3e}} \!& \simeq &\!  N_\mu \,R_\mu \,\delta t \,P_\text{p}\, B_{\mu^+ \rightarrow e^+ \nu_e \bar\nu_\mu}\, B_{\mu^+ \rightarrow e^+ \nu_e\bar\nu_\mu  e^+e^-}\,,~~
\label{eq:2M3e}
\eea
where $N_\mu = 10^{15}$ ($2\times 10^{16}$) is the total number of muons and $R_\mu=10^8$/s ($2 \times 10^9$/s) is the instantaneous 
stopped muon rate for Mu3e phase~I~(II), $\delta t=2.5\times 10^{-10}\rm \, s$ is the average time resolution, $P_\text{p} = 10^{-4}$ is the position 
suppression factor, $P_{e^+\to e^-}=0.5\%$ is the positron-to-electron misidentification probability, and $P_\gamma = 8 \times 10^{-4}$ is 
the photon conversion probability in the target. 
Derivations of~\eqssref{3M}{2My}{2M3e} are shown in Appendix~\ref{sec:pu}.  
Inserting the numbers, we find that the expected number of accidental background events over the lifetime of the experiment 
(before correcting for the efficiency) are given by, roughly,  
$N_{\rm 3 M} \sim 15,000$ (60,000), $N_{\rm 2 M_\gamma} \sim$~30,000 ($6\times 10^6$), and 
$N_{\rm 2 M_e} \sim$~75,000 ($2\times 10^7$) for phase~I~(II).     
We use these numbers to normalize each accidental background component.
We note that we will not consider the 2M$_{\rm Bhabha}$ background, as it is challenging to simulate reliably.  
More study is needed by the \mue\ Collaboration to determine its size, but preliminary 
estimates\footnote{We thank Andr\'e Sch\"oning for valuable discussions of this background.} suggest that 
in the 10~MeV to 80~MeV mass range, this background should be at most comparable, but more likely 
subdominant, to the irreducible backgrounds.   

The $e^+e^-$ invariant mass distribution of the most important irreducible and accidental backgrounds, 
after applying all selection criteria, is shown in Fig.~\ref{Fig::eemass}, assuming a total number  of $10^{15}$ (top plot) 
and $5.5\times 10^{16}$ (bottom plot) muon decays for the two phases of Mu3e, respectively.  
Both combinations per muon candidate are considered and included in the corresponding histograms. 
The signal reconstruction efficiency is shown in Fig.~\ref{fig:effiPrompt} and varies between 7\% and 41\%, 
depending on the dark photon mass.  

As expected, the distribution peaks towards low values of $m_{e^+e^-}$. 
The spectrum is dominated by $\mu^+ \rightarrow e^+ \nu_e\bar\nu_\mu  e^+e^-$ 
events (red line in Fig.~\ref{Fig::eemass}) with an additional contribution 
from $\mu^+ \rightarrow e^+\nu_e\bar\nu_\mu \gamma$ with the conversion 
$\gamma \rightarrow \epem$ in the target material (black line).  
The accidental backgrounds (green line) are subdominant, except for $m_{e^+ e^-} \gtrsim 80$~MeV, 
where they become comparable to the irreducible contribution. 
However, as we will discuss below, this region is already well explored by existing experiments.  
Therefore, even if our accidental background estimate is off by a factor of a few, it will have little impact on the 
the dark-photon parameter region probed by Mu3e that is currently unexplored ($m_{A'} \lesssim 70 \mev$).  

\begin{figure}[!t]
\begin{center}
\includegraphics[width=0.45\textwidth]{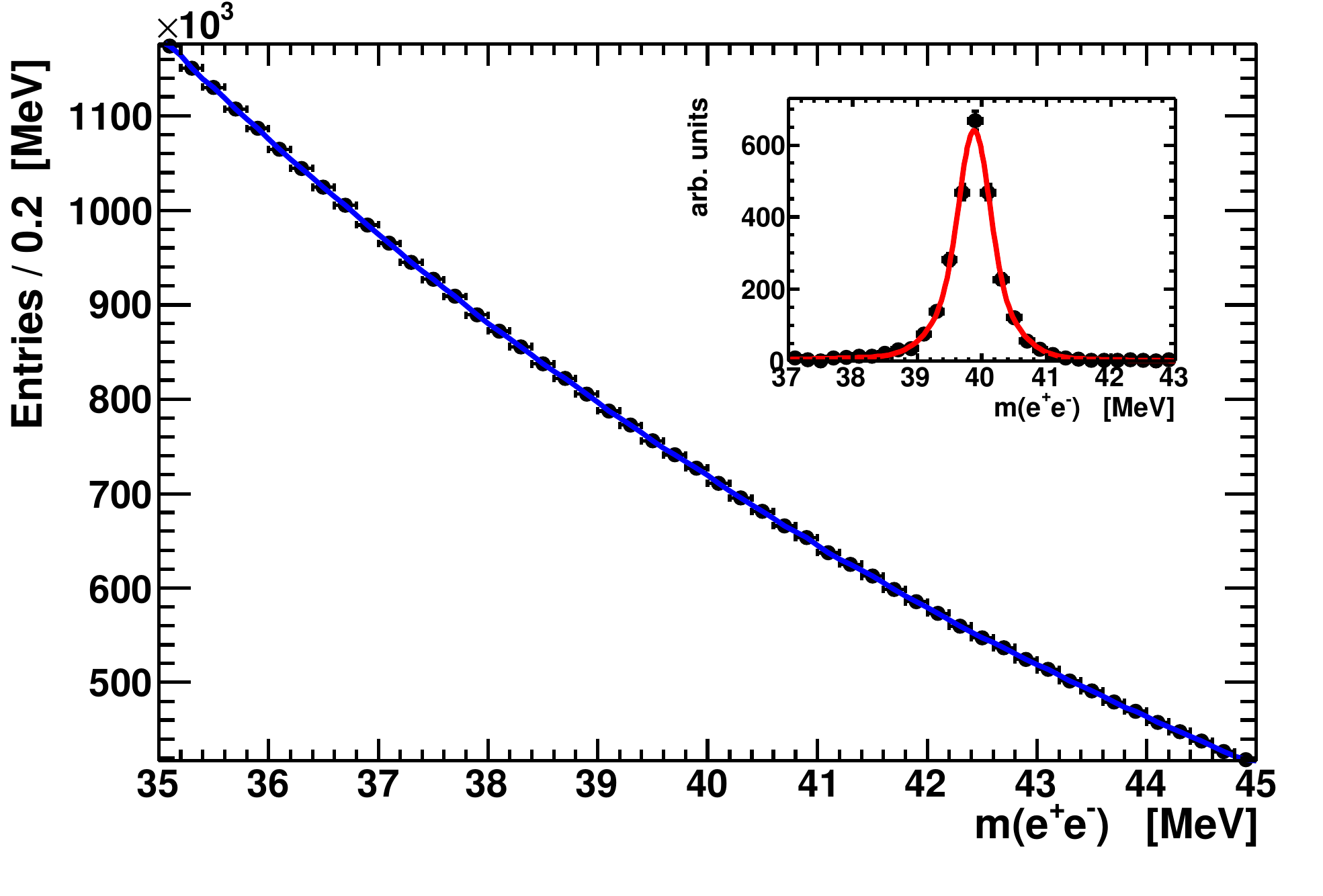}
\end{center}
\caption{Fit to the $\epem$ invariant mass distribution for a dark photon mass hypothesis of $40 \mev$. The blue line shows the 
expected background. The signal probability density function, as obtained from a fit to the signal Monte Carlo sample, is 
shown as the red line in the insert. 
}
\label{Fig::Fit30}
\end{figure}
\begin{figure*}[!t]
    \centering
    \includegraphics[width=0.75 \textwidth]{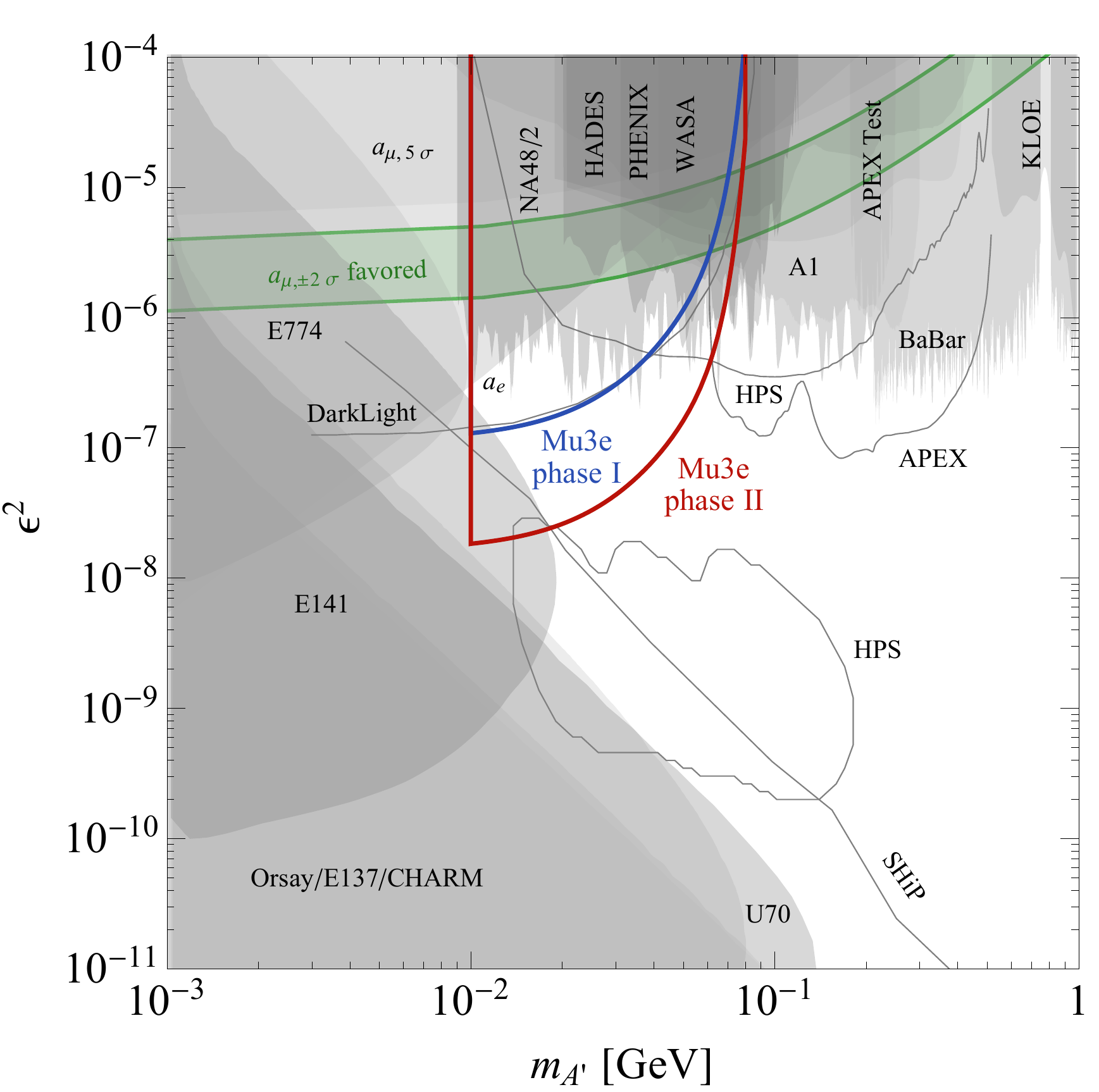}
        \caption{Prospects and constraints in the $\epsilon^2$ versus $m_{A'}$ plane for dark photons that decay 
        directly to SM particles (see \eg~\cite{Essig:2013lka} and references given in Sec.~\ref{sec:intro} of this paper).  
        The projected sensitivity of a resonance search for promptly decaying dark photons with the Mu3e experiment is 
        shown in blue (red) assuming $10^{15}$ ($5.5\times 10^{16}$) muon decays for Mue3 phase I (II).  
      }
    \label{fig:A'-visible}
 \end{figure*}

A dark photon signal would appear as a narrow peak over the smooth background distribution. 
The signal resolution is determined by fitting the corresponding mass spectrum with a sum of three Gaussians. 
The central mass resolution is at the level of $0.2-0.3 \mev$, almost independent of $m_{A'}$. 
We checked that these results are similar to the expected performance of the Mu3e detector~\cite{Blondel:2013ia}.

We estimate the signal sensitivity by fitting a signal component on top of the expected background in 
the range $10\mev < m_{A'} < 80 \mev$. Each fit is performed over an interval of $\pm 5 \mev$ around the nominal dark 
photon mass.. An example of a fit is displayed in Fig.~\ref{Fig::Fit30}. 
We extract a 95\% confidence level (CL) limit on the number of signal events, and derive a bound on the 
$\mu^+ \rightarrow e^+ \nu_e \bar\nu_\mu A',~A' \rightarrow e^+e^-$ branching fraction by dividing by the signal efficiency and 
the number of muon decays. These results are translated into limits on the kinetic mixing parameter, $\epsilon$, and 
shown as a blue (red) solid line for Mu3e's phase~I (II) in Fig.~\ref{fig:A'-visible}, together with existing constraints 
and prospects for upcoming experiments. 

A substantial fraction of open parameter space in the low $m_{A'}$ region can be explored, complementing or overlapping 
the reach of currently planned experiments, including APEX~\cite{Essig:2010xa,Abrahamyan:2011gv}, HPS~\cite{Battaglieri:2014hga}, 
DarkLight~\cite{Freytsis:2009bh,Balewski:2013oza}, and an experiment at the SPS~\cite{Andreas:2013lya} 
(the latter is not shown). 
As mentioned in the introduction, if U(1)$_\text{Y}$ is embedded in a GUT, the mixing that is generated by a one-(two-)loop interaction 
naturally gives $\epsilon^2 \sim 10^{-6}-10^{-2}$ ($\sim 10^{-10}-10^{-6}$).  
Mu3e has the opportunity to explore part of this theoretically interesting parameter space.

\subsection{Displaced Vertices From Dark Photons}
\label{sec:vertexregion}

For sufficiently small values of $\epsilon$, the dark photon lifetime can be sizable (see Eq.~(\ref{eq:width})), 
leading to displaced decay vertices observable in the laboratory frame.
While smaller values of $\epsilon$ lead to smaller muon branching fractions to dark photons, the backgrounds associated with 
displaced vertices are substantially reduced, providing an opportunity to observe a signal. 
The discovery potential depends on the geometrical acceptance of the detector, the vertex resolution, and the backgrounds. 
The assumptions introduced in the prompt decay scenario to treat the accidental backgrounds, \ie\ pile-up events arising from 
muons decaying within the same time window and at the same position in the target, 
might not be valid anymore for displaced vertices. 
A full analysis should include contributions from pile-up of several (radiative) muon decays generated everywhere in the target, 
which is  beyond the scope of our paper, given the large number of muon decays involved. Furthermore, a small 
residual background from misreconstructed $\mu^+ \rightarrow e^+ \nu_e\bar\nu_\mu  e^+e^-$ events is expected to remain, 
and the accuracy of FastSim might be too limited to reliably predict its level. 
We encourage the \mue\ Collaboration to perform a detailed reach estimate, both because the tools 
at our disposal are not sufficient for a reliable estimate and because the sensitivity that could potentially be achieved is well worth 
the effort.

\section{Conclusions}
\label{sec:conclusions}

We have studied the possibility to search for dark photon in $\mu^+ \rightarrow e^+ \nu_e \bar\nu_\mu A',~A' \rightarrow e^+e^-$ 
decays with an apparatus similar to the Mu3e experiment. 
We derive sensitivity estimates for both prompt and displaced dark photon decays. 
Mu3e has the exciting opportunity to probe a substantial fraction of currently unexplored parameter space in the mass range 
$10\mev \lesssim m_{A'} \lesssim 80 \mev$ for $\epsilon^2 \gtrsim 10^{-8}$, using a resonance search, 
overlapping or complementing the reach of currently planned experiments. 
This opportunity does not require any modifications of their existing setup.  
A search for displaced vertices may have sensitivity to lower values of $\epsilon$, but the precise reach estimate depends on the 
backgrounds, which require a careful modeling by the Mu3e Collaboration.

\subsection*{Acknowledgements}
We thank Andr\'e Sch\"oning for reading and providing useful comments on the manuscript.  
BE is supported by the U.S.~Department of Energy (DoE) under grant DE-FG02-92ER40701 and DE-SC0011925.
RE and YZ are supported by the DoE Early Career research program DESC0008061.  
RE acknowledges additional support through a Sloan Foundation Research Fellowship.

\appendix
\section{Estimation of accidental backgrounds}
\label{sec:pu}
In Sec.~\ref{sec:exclusionregion}, we provided an estimate of the accidental backgrounds, see~\eqssref{3M}{2My}{2M3e}.  
In this appendix, we justify these equations.  

We first estimate the accidental background arising from three Michel decays, where one positron is misreconstructed as an 
electron ($N_{\rm 3M}$).  We assume a measurement takes place in a time interval $T \gg \delta t$, where $\delta t$ is the 
time resolution.  The probability for three decays to occur in the same time window is $(\delta t /T)^2$ and 
at the same position is $P_{\text p}^2$, where $P_{\text p}$ is the position suppression factor.  
Multiplying these probabilities with the branching ratio for three Michel decays, $B_{\mu^+ \rightarrow e^+ \nu_e \bar\nu_\mu}^3$, 
and the probability for one out of three positrons to be misreconstructed as an electron, $P_{e^+\to e^-}$, 
the total probability for the 3M pile-up is given by
\beq
P_{\text{3M}}=\left(\frac{\delta t}{T} \right)^2 \,P_{\text p}^2\, B_{\mu^+ \rightarrow e^+ \nu_e \bar\nu_\mu}^3 \,   \begin{pmatrix}    
   3 \\
      1 \\
   \end{pmatrix}\,P_{e^+\to e^-}.
\label{eq:totprob}
\eeq
During the time $T$, the number of stopped muons, $N_\mu\equiv R_\mu T$, and the number of 3M pile-up event, $N_{\text{3M}}$, are related by
\beq
N_{\text{3M}}=\begin{pmatrix} 
      N_\mu \\
      3 \\
   \end{pmatrix} P_{\text{3M}}.
   \label{eq:totN}
\eeq
Substituting \Eqref{totprob} into \Eqref{totN} yields
\bea
N_{\text{3M}}&=&\frac{1}{2}T R_\mu \left(R_\mu -\frac{1}{T}\right) \left( R_\mu -\frac{2}{T}\right) \delta t^2 P_{\text p}^2 B_{\mu^+ \rightarrow e^+ \nu_e \bar\nu_\mu}^3 \nonumber \\
&\simeq& \frac{1}{2} N_\mu \,R_{\mu}^2\, \delta t^2 \,P_{\text p}^2 \,B_{\mu^+ \rightarrow e^+ \nu_e \bar\nu_\mu}^3\,,
\eea
which is Eq.~(\ref{eq:3M}).  The approximation is valid in the limit $R_\mu T \gg 1$. 

Similarly, the total probabilities for the 2M$_\gamma$ and 2M$_{3e}$ accidental backgrounds are given by 
\bea
 P_{\text{2M}_\gamma}&=&\begin{pmatrix}2 \\ 1\end{pmatrix} \,B_{\mu^+ \rightarrow e^+ \nu_e \bar\nu_\mu}\,  B_{\mu^+ \rightarrow e^+ \nu_e \bar\nu_\mu \gamma}  \,P_\gamma,\\
 P_{\text{2M}_{3e}}&=&\begin{pmatrix}2 \\ 1\end{pmatrix} B_{\mu^+ \rightarrow e^+ \nu_e \bar\nu_\mu}\, B_{\mu^+ \rightarrow e^+ \nu_e\bar\nu_\mu  e^+e^-},
\eea
respectively. 
Given the number of 2M$_\gamma$ and 2M$_{3 e}$ as 
\bea
N_{\text{2M}_{\gamma}} &=& \begin{pmatrix} 
      N_\mu \\
      2 \\
   \end{pmatrix} P_{\text{2M}_{\gamma}}\,, \\
N_{\text{2M}_{3e}} &=& \begin{pmatrix} 
      N_\mu \\
      2 \\
   \end{pmatrix} P_{\text{2M}_{3e}}\,,
\eea
one finds 
\bea
N_{\text{2M}_\gamma}\!&=&\!T R_\mu \left(R_\mu -\frac{1}{T}\right) \delta t P_{\text p} B_{\mu^+ \rightarrow e^+ \nu_e \bar\nu_\mu}\,  B_{\mu^+ \rightarrow e^+ \nu_e \bar\nu_\mu \gamma}  \,P_\gamma \nonumber\\
\!&\simeq&\! T R_\mu ^2 \delta t P_\text{p} B_{\mu^+ \rightarrow e^+ \nu_e \bar\nu_\mu}\,  B_{\mu^+ \rightarrow e^+ \nu_e \bar\nu_\mu \gamma}  \,P_\gamma\,, \\
N_{\text{2M}_{3e}}\!&=&\!T R_\mu \left(R_\mu -\frac{1}{T}\right) \delta t P_{\text p}  B_{\mu^+ \rightarrow e^+ \nu_e \bar\nu_\mu}\, B_{\mu^+ \rightarrow e^+ \nu_e\bar\nu_\mu  e^+e^-} \nonumber\\
\!&\simeq&\! T R_\mu ^2 \delta t P_\text{p} B_{\mu^+ \rightarrow e^+ \nu_e \bar\nu_\mu}\, B_{\mu^+ \rightarrow e^+ \nu_e\bar\nu_\mu  e^+e^-}.
\eea
These are Eqs.~(\ref{eq:2My}) and (\ref{eq:2M3e}).  
The approximations in the above equations are again valid for $R_\mu T \gg 1$.

\bibliography{mu3e}

\providecommand{\href}[2]{#2}\begingroup\raggedright\begin{thebibliography}{10}

\bibitem{Essig:2013lka}
R.~Essig, J.~A. Jaros, W.~Wester, et~al., {\it {Dark Sectors and New, Light,
  Weakly-Coupled Particles}},  \href{http://xxx.lanl.gov/abs/1311.0029}{{\tt
  arXiv:1311.0029}}.

\bibitem{Hewett:2012ns}
J.~Hewett, H.~Weerts, R.~Brock, J.~Butler, B.~Casey, et~al., {\it {Fundamental
  Physics at the Intensity Frontier}},
  \href{http://xxx.lanl.gov/abs/1205.2671}{{\tt arXiv:1205.2671}}.

\bibitem{Jaeckel:2010ni}
J.~Jaeckel and A.~Ringwald, {\it {The Low-Energy Frontier of Particle
  Physics}},  {\em Ann.Rev.Nucl.Part.Sci.} {\bf 60} (2010) 405--437,
  [\href{http://xxx.lanl.gov/abs/1002.0329}{{\tt arXiv:1002.0329}}].

\bibitem{Holdom:1985ag}
B.~Holdom, {\it {Two U(1)'s and Epsilon Charge Shifts}},  {\em Phys.Lett.} {\bf
  B166} (1986) 196.

\bibitem{Galison:1983pa}
P.~Galison and A.~Manohar, {\it {Two Z's or Not Two Z's?}},  {\em Phys.Lett.}
  {\bf B136} (1984) 279.

\bibitem{Dienes:1996zr}
K.~R. Dienes, C.~F. Kolda, and J.~March-Russell, {\it {Kinetic mixing and the
  supersymmetric gauge hierarchy}},  {\em Nucl.Phys.} {\bf B492} (1997)
  104--118, [\href{http://xxx.lanl.gov/abs/hep-ph/9610479}{{\tt
  hep-ph/9610479}}].

\bibitem{Pospelov:2008zw}
M.~Pospelov, {\it {Secluded U(1) below the weak scale}},  {\em Phys. Rev.} {\bf
  D80} (2009) 095002.

\bibitem{Bennett:2006fi}
{\bf Muon G-2 Collaboration} Collaboration, G.~Bennett et~al., {\it {Final
  Report of the Muon E821 Anomalous Magnetic Moment Measurement at BNL}},  {\em
  Phys.Rev.} {\bf D73} (2006) 072003,
  [\href{http://xxx.lanl.gov/abs/hep-ex/0602035}{{\tt hep-ex/0602035}}].

\bibitem{Davier:2010nc}
M.~Davier, A.~Hoecker, B.~Malaescu, and Z.~Zhang, {\it {Reevaluation of the
  Hadronic Contributions to the Muon g-2 and to alpha(MZ)}},  {\em Eur.Phys.J.}
  {\bf C71} (2011) 1515, [\href{http://xxx.lanl.gov/abs/1010.4180}{{\tt
  arXiv:1010.4180}}].

\bibitem{ArkaniHamed:2008qn}
N.~Arkani-Hamed, D.~P. Finkbeiner, T.~R. Slatyer, and N.~Weiner, {\it {A Theory
  of Dark Matter}},  {\em Phys.Rev.} {\bf D79} (2009) 015014,
  [\href{http://xxx.lanl.gov/abs/0810.0713}{{\tt arXiv:0810.0713}}].

\bibitem{Pospelov:2008jd}
M.~Pospelov and A.~Ritz, {\it {Astrophysical Signatures of Secluded Dark
  Matter}},  {\em Phys.Lett.} {\bf B671} (2009) 391--397,
  [\href{http://xxx.lanl.gov/abs/0810.1502}{{\tt arXiv:0810.1502}}].

\bibitem{Finkbeiner:2007kk}
D.~P. Finkbeiner and N.~Weiner, {\it {Exciting Dark Matter and the INTEGRAL/SPI
  511 keV signal}},  {\em Phys.Rev.} {\bf D76} (2007) 083519,
  [\href{http://xxx.lanl.gov/abs/astro-ph/0702587}{{\tt astro-ph/0702587}}].

\bibitem{Fayet:2004bw}
P.~Fayet, {\it {Light spin 1/2 or spin 0 dark matter particles}},  {\em
  Phys.Rev.} {\bf D70} (2004) 023514,
  [\href{http://xxx.lanl.gov/abs/hep-ph/0403226}{{\tt hep-ph/0403226}}].

\bibitem{ArkaniHamed:2008qp}
N.~Arkani-Hamed and N.~Weiner, {\it {LHC Signals for a SuperUnified Theory of
  Dark Matter}},  {\em JHEP} {\bf 0812} (2008) 104,
  [\href{http://xxx.lanl.gov/abs/0810.0714}{{\tt arXiv:0810.0714}}].

\bibitem{Cheung:2009fk}
C.~Cheung, J.~T. Ruderman, L.-T. Wang, and I.~Yavin, {\it {Kinetic Mixing as
  the Origin of Light Dark Scales}},
  \href{http://xxx.lanl.gov/abs/0902.3246}{{\tt arXiv:0902.3246}}.

\bibitem{Baumgart:2009tn}
M.~Baumgart, C.~Cheung, J.~T. Ruderman, L.-T. Wang, and I.~Yavin, {\it
  {Non-Abelian Dark Sectors and Their Collider Signatures}},  {\em JHEP} {\bf
  0904} (2009) 014, [\href{http://xxx.lanl.gov/abs/0901.0283}{{\tt
  arXiv:0901.0283}}].

\bibitem{Morrissey:2009ur}
D.~E. Morrissey, D.~Poland, and K.~M. Zurek, {\it {Abelian Hidden Sectors at a
  GeV}},  {\em JHEP} {\bf 0907} (2009) 050,
  [\href{http://xxx.lanl.gov/abs/0904.2567}{{\tt arXiv:0904.2567}}].

\bibitem{Essig:2009nc}
R.~Essig, P.~Schuster, and N.~Toro, {\it {Probing Dark Forces and Light Hidden
  Sectors at Low-Energy e+e- Colliders}},  {\em Phys. Rev.} {\bf D80} (2009)
  015003, [\href{http://xxx.lanl.gov/abs/0903.3941}{{\tt arXiv:0903.3941}}].

\bibitem{Essig:2010ye}
R.~Essig, J.~Kaplan, P.~Schuster, and N.~Toro, {\it {On the Origin of Light
  Dark Matter Species}},  {\em Submitted to Physical Review D} (2010)
  [\href{http://xxx.lanl.gov/abs/1004.0691}{{\tt arXiv:1004.0691}}].

\bibitem{Bjorken:2009mm}
J.~D. Bjorken, R.~Essig, P.~Schuster, and N.~Toro, {\it {New Fixed-Target
  Experiments to Search for Dark Gauge Forces}},  {\em Phys. Rev.} {\bf D80}
  (2009) 075018.

\bibitem{Bjorken:1988as}
J.~D. Bjorken et~al., {\it {Search for Neutral Metastable Penetrating Particles
  Produced in the SLAC Beam Dump}},  {\em Phys. Rev.} {\bf D38} (1988) 3375.

\bibitem{Riordan:1987aw}
E.~M. Riordan et~al., {\it {A Search for Short Lived Axions in an Electron Beam
  Dump Experiment}},  {\em Phys. Rev. Lett.} {\bf 59} (1987) 755.

\bibitem{Bross:1989mp}
A.~Bross et~al., {\it {A Search for Shortlived Particles Produced in an
  Electron Beam Dump}},  {\em Phys. Rev. Lett.} {\bf 67} (1991) 2942--2945.

\bibitem{Batell:2009yf}
B.~Batell, M.~Pospelov, and A.~Ritz, {\it {Probing a Secluded U(1) at
  B-factories}},  {\em Phys. Rev.} {\bf D79} (2009) 115008.

\bibitem{Strassler:2006im}
M.~J. Strassler and K.~M. Zurek, {\it {Echoes of a hidden valley at hadron
  colliders}},  {\em Phys.Lett.} {\bf B651} (2007) 374--379,
  [\href{http://xxx.lanl.gov/abs/hep-ph/0604261}{{\tt hep-ph/0604261}}].

\bibitem{Strassler:2006qa}
M.~J. Strassler, {\it {Possible effects of a hidden valley on supersymmetric
  phenomenology}},  \href{http://xxx.lanl.gov/abs/hep-ph/0607160}{{\tt
  hep-ph/0607160}}.

\bibitem{Freytsis:2009bh}
M.~Freytsis, G.~Ovanesyan, and J.~Thaler, {\it {Dark Force Detection in Low
  Energy e-p Collisions}},  {\em JHEP} {\bf 1001} (2010) 111,
  [\href{http://xxx.lanl.gov/abs/0909.2862}{{\tt arXiv:0909.2862}}].

\bibitem{Essig:2010xa}
R.~Essig, P.~Schuster, N.~Toro, and B.~Wojtsekhowski, {\it {An Electron Fixed
  Target Experiment to Search for a New Vector Boson A' Decaying to e+e-}},
  {\em JHEP} {\bf 1102} (2011) 009,
  [\href{http://xxx.lanl.gov/abs/1001.2557}{{\tt arXiv:1001.2557}}].

\bibitem{Blumlein:2011mv}
J.~Blumlein and J.~Brunner, {\it {New Exclusion Limits for Dark Gauge Forces
  from Beam-Dump Data}},  {\em Phys.Lett.} {\bf B701} (2011) 155--159,
  [\href{http://xxx.lanl.gov/abs/1104.2747}{{\tt arXiv:1104.2747}}].

\bibitem{Andreas:2012mt}
S.~Andreas, C.~Niebuhr, and A.~Ringwald, {\it {New Limits on Hidden Photons
  from Past Electron Beam Dumps}},  {\em Phys.Rev.} {\bf D86} (2012) 095019,
  [\href{http://xxx.lanl.gov/abs/1209.6083}{{\tt arXiv:1209.6083}}].

\bibitem{Reece:2009un}
M.~Reece and L.-T. Wang, {\it {Searching for the light dark gauge boson in
  GeV-scale experiments}},  {\em JHEP} {\bf 07} (2009) 051.

\bibitem{Aubert:2009cp}
{\bf BaBar Collaboration} Collaboration, B.~Aubert et~al., {\it {Search for
  Dimuon Decays of a Light Scalar Boson in Radiative Transitions $\Upsilon
  \rightarrow \gamma A^0$}},  {\em Phys.Rev.Lett.} {\bf 103} (2009) 081803,
  [\href{http://xxx.lanl.gov/abs/0905.4539}{{\tt arXiv:0905.4539}}].

\bibitem{Hook:2010tw}
A.~Hook, E.~Izaguirre, and J.~G. Wacker, {\it {Model Independent Bounds on
  Kinetic Mixing}},  {\em Adv.High Energy Phys.} {\bf 2011} (2011) 859762,
  [\href{http://xxx.lanl.gov/abs/1006.0973}{{\tt arXiv:1006.0973}}].

\bibitem{Babusci:2012cr}
{\bf KLOE-2 Collaboration} Collaboration, D.~Babusci et~al., {\it {Limit on the
  production of a light vector gauge boson in phi meson decays with the KLOE
  detector}},  {\em Phys.Lett.} {\bf B720} (2013) 111--115,
  [\href{http://xxx.lanl.gov/abs/1210.3927}{{\tt arXiv:1210.3927}}].

\bibitem{Archilli:2011zc}
F.~Archilli, D.~Babusci, D.~Badoni, I.~Balwierz, G.~Bencivenni, et~al., {\it
  {Search for a vector gauge boson in phi meson decays with the KLOE
  detector}},  {\em Phys.Lett.} {\bf B706} (2012) 251--255,
  [\href{http://xxx.lanl.gov/abs/1110.0411}{{\tt arXiv:1110.0411}}].

\bibitem{Abrahamyan:2011gv}
{\bf APEX} Collaboration, S.~Abrahamyan et~al., {\it {Search for a new gauge
  boson in the $A'$ Experiment (APEX)}},  {\em Phys. Rev. Lett.} {\bf 107}
  (2011) 191804, [\href{http://xxx.lanl.gov/abs/1108.2750}{{\tt
  arXiv:1108.2750}}].

\bibitem{Merkel:2014avp}
H.~Merkel, P.~Achenbach, C.~A. Gayoso, T.~Beranek, J.~Bericic, et~al., {\it
  {Search for light massive gauge bosons as an explanation of the $(g-2)_\mu$
  anomaly at MAMI}},  \href{http://xxx.lanl.gov/abs/1404.5502}{{\tt
  arXiv:1404.5502}}.

\bibitem{Dent:2012mx}
J.~B. Dent, F.~Ferrer, and L.~M. Krauss, {\it {Constraints on Light Hidden
  Sector Gauge Bosons from Supernova Cooling}},
  \href{http://xxx.lanl.gov/abs/1201.2683}{{\tt arXiv:1201.2683}}.

\bibitem{Davoudiasl:2012ig}
H.~Davoudiasl, H.-S. Lee, and W.~J. Marciano, {\it {Dark Side of Higgs Diphoton
  Decays and Muon g-2}},  {\em Phys.Rev.} {\bf D86} (2012) 095009,
  [\href{http://xxx.lanl.gov/abs/1208.2973}{{\tt arXiv:1208.2973}}].

\bibitem{Davoudiasl:2012ag}
H.~Davoudiasl, H.-S. Lee, and W.~J. Marciano, {\it {'Dark' Z implications for
  Parity Violation, Rare Meson Decays, and Higgs Physics}},  {\em Phys.Rev.}
  {\bf D85} (2012) 115019, [\href{http://xxx.lanl.gov/abs/1203.2947}{{\tt
  arXiv:1203.2947}}].

\bibitem{Davoudiasl:2013aya}
H.~Davoudiasl, H.-S. Lee, I.~Lewis, and W.~J. Marciano, {\it {Higgs Decays as a
  Window into the Dark Sector}},  \href{http://xxx.lanl.gov/abs/1304.4935}{{\tt
  arXiv:1304.4935}}.

\bibitem{Endo:2012hp}
M.~Endo, K.~Hamaguchi, and G.~Mishima, {\it {Constraints on Hidden Photon
  Models from Electron g-2 and Hydrogen Spectroscopy}},  {\em Phys.Rev.} {\bf
  D86} (2012) 095029, [\href{http://xxx.lanl.gov/abs/1209.2558}{{\tt
  arXiv:1209.2558}}].

\bibitem{Balewski:2013oza}
J.~Balewski, J.~Bernauer, W.~Bertozzi, J.~Bessuille, B.~Buck, et~al., {\it
  {DarkLight: A Search for Dark Forces at the Jefferson Laboratory
  Free-Electron Laser Facility}},
  \href{http://xxx.lanl.gov/abs/1307.4432}{{\tt arXiv:1307.4432}}.

\bibitem{Adlarson:2013eza}
{\bf WASA-at-COSY Collaboration} Collaboration, P.~Adlarson et~al., {\it
  {Search for a dark photon in the $\pi^0 \to e^+e^-\gamma$ decay}},  {\em
  Phys.Lett.} {\bf B726} (2013) 187--193,
  [\href{http://xxx.lanl.gov/abs/1304.0671}{{\tt arXiv:1304.0671}}].

\bibitem{Agakishiev:2013fwl}
{\bf HADES} Collaboration, G.~Agakishiev et~al., {\it {Searching a Dark Photon
  with HADES}},  {\em Phys.Lett.} {\bf B731} (2014) 265--271,
  [\href{http://xxx.lanl.gov/abs/1311.0216}{{\tt arXiv:1311.0216}}].

\bibitem{Andreas:2013lya}
S.~Andreas, S.~Donskov, P.~Crivelli, A.~Gardikiotis, S.~Gninenko, et~al., {\it
  {Proposal for an Experiment to Search for Light Dark Matter at the SPS}},
  \href{http://xxx.lanl.gov/abs/1312.3309}{{\tt arXiv:1312.3309}}.

\bibitem{Battaglieri:2014hga}
M.~Battaglieri, S.~Boyarinov, S.~Bueltmann, V.~Burkert, A.~Celentano, et~al.,
  {\it {The Heavy Photon Search Test Detector}},
  \href{http://xxx.lanl.gov/abs/1406.6115}{{\tt arXiv:1406.6115}}.

\bibitem{Lees:2014xha}
{\bf BaBar} Collaboration, J.~Lees et~al., {\it {Search for a dark photon in
  e+e- collisions at BABAR}},  \href{http://xxx.lanl.gov/abs/1406.2980}{{\tt
  arXiv:1406.2980}}.

\bibitem{Adare:2014ega}
A.~Adare, S.~Afanasiev, C.~Aidala, N.~Ajitanand, Y.~Akiba, et~al., {\it
  {Closing the Door for Dark Photons as the Explanation for the Muon g-2
  Anomaly}},  \href{http://xxx.lanl.gov/abs/1409.0851}{{\tt arXiv:1409.0851}}.

\bibitem{Kazanas:2014mca}
D.~Kazanas, R.~N. Mohapatra, S.~Nussinov, V.~Teplitz, and Y.~Zhang, {\it
  {Supernova Bounds on the Dark Photon Using its Electromagnetic Decay}},
  \href{http://xxx.lanl.gov/abs/1410.0221}{{\tt arXiv:1410.0221}}.

\bibitem{Blumlein:2013cua}
J.~Bl{\"u}mlein and J.~Brunner, {\it {New Exclusion Limits on Dark Gauge Forces
  from Proton Bremsstrahlung in Beam-Dump Data}},  {\em Phys.Lett.} {\bf B731}
  (2014) 320--326, [\href{http://xxx.lanl.gov/abs/1311.3870}{{\tt
  arXiv:1311.3870}}].

\bibitem{CERNNA48/2:2015lha}
{\bf CERN NA48/2} Collaboration, C.~N. Collaboration, {\it {Search for the dark
  photon in $\pi^0$ decays}},  \href{http://xxx.lanl.gov/abs/1504.0060}{{\tt
  arXiv:1504.0060}}.

\bibitem{Blondel:2013ia}
A.~Blondel, A.~Bravar, M.~Pohl, S.~Bachmann, N.~Berger, et~al., {\it {Research
  Proposal for an Experiment to Search for the Decay $\mu \to eee$}},
  \href{http://xxx.lanl.gov/abs/1301.6113}{{\tt arXiv:1301.6113}}.

\bibitem{Djilkibaev:2008jy}
R.~M. Djilkibaev and R.~V. Konoplich, {\it {Rare Muon Decay $\mu^+ \to e^+ e^-
  e^+ \nu_e \bar\nu_\mu$}},  {\em Phys.Rev.} {\bf D79} (2009) 073004,
  [\href{http://xxx.lanl.gov/abs/0812.1355}{{\tt arXiv:0812.1355}}].

\bibitem{Alwall:2014hca}
J.~Alwall, R.~Frederix, S.~Frixione, V.~Hirschi, F.~Maltoni, et~al., {\it {The
  automated computation of tree-level and next-to-leading order differential
  cross sections, and their matching to parton shower simulations}},  {\em
  JHEP} {\bf 1407} (2014) 079, [\href{http://xxx.lanl.gov/abs/1405.0301}{{\tt
  arXiv:1405.0301}}].

\bibitem{pdg}
{\bf Particle Data Group} Collaboration, K.~Olive et~al., {\it {Review of
  Particle Physics (RPP)}},  {\em Chin.Phys.} {\bf C38} (2014) 090001.

\bibitem{Baszczyk:2013xua}
{\bf SuperB Collaboration} Collaboration, M.~Baszczyk et~al., {\it {SuperB
  Technical Design Report}},  \href{http://xxx.lanl.gov/abs/1306.5655}{{\tt
  arXiv:1306.5655}}.

\bibitem{Aubert:2001tu}
{\bf BaBar Collaboration} Collaboration, B.~Aubert et~al., {\it {The BaBar
  detector}},  {\em Nucl.Instrum.Meth.} {\bf A479} (2002) 1--116,
  [\href{http://xxx.lanl.gov/abs/hep-ex/0105044}{{\tt hep-ex/0105044}}].

\bibitem{TheBABAR:2013jta}
{\bf BABAR} Collaboration, B.~Aubert et~al., {\it {The BABAR Detector:
  Upgrades, Operation and Performance}},  {\em Nucl.Instrum.Meth.} {\bf A729}
  (2013) 615--701, [\href{http://xxx.lanl.gov/abs/1305.3560}{{\tt
  arXiv:1305.3560}}].

\bibitem{Kuno:1999jp}
Y.~Kuno and Y.~Okada, {\it {Muon decay and physics beyond the standard model}},
   {\em Rev.Mod.Phys.} {\bf 73} (2001) 151--202,
  [\href{http://xxx.lanl.gov/abs/hep-ph/9909265}{{\tt hep-ph/9909265}}].

\end{thebibliography}\endgroup
\bibliographystyle{JHEP}

\end{document}